\documentclass[12pt]{article}
\usepackage{amsmath}
\usepackage{graphicx,psfrag,epsf}
\usepackage{enumerate}
\usepackage{natbib}
\usepackage{booktabs,caption}
\usepackage[flushleft]{threeparttable}
\usepackage[capposition=top]{floatrow}
\usepackage{url} 
\usepackage[normalem]{ulem}
\useunder{\uline}{\ul}{}

\newcommand{\blind}{0}

\defcitealias{eeoc}{USEEOC, 2016}
\defcitealias{uscacs}{Census Bureau, 2014}

\begin{document}

\def\spacingset#1{\renewcommand{\baselinestretch}%
{#1}\small\normalsize} \spacingset{1}


\if0\blind
{
  \title{\bf Better estimates from binned income data: Interpolated CDFs and mean-matching}

  \author{    Paul T. von Hippel\\
    LBJ School of Public Affairs, University of Texas at Austin \\ paulvonhippel.utaustin@gmail.com \\
           \\
          David J. Hunter, McKalie Drown\thanks{
    Drown is grateful for support from a Tensor Grant of the Mathematical Association of America.}\hspace{.2cm} \\
    Department of Mathematics and Computer Science, Westmont College \\
    dhunter@westmont.edu, mdrown@westmont.edu
    }

  \maketitle
} \fi

\if1\blind
{
  \bigskip
  \bigskip
  \bigskip
  \begin{center}
    {\Large\bf Better estimates from binned incomes: 
    Interpolated CDFs and mean-matching}
\end{center}
  \medskip
} \fi

\bigskip
\begin{abstract}
Researchers often estimate income statistics from summaries that report the number of incomes in bins such as \$0-10,000, \$10,001-20,000,\ldots,\$200,000+. Some analysts assign incomes to bin midpoints, but this treats income as discrete. Other analysts fit a continuous parametric distribution, but the distribution may not fit well. 

We fit nonparametric continuous distributions that reproduce the bin counts perfectly by interpolating the cumulative distribution function (CDF). We also show how both midpoints and interpolated CDFs can be constrained to reproduce the mean of income when it is known.

We evaluate the methods in estimating the Gini coefficients of all 3,221 US counties. Fitting parametric distributions is very slow. Fitting interpolated CDFs is much faster and slightly more accurate. Both interpolated CDFs and midpoints give dramatically better estimates if constrained to match a known mean.

We have implemented interpolated CDFs in the \textit{binsmooth} package for R. We have implemented the midpoint method in the \textit{rpme} command for Stata. Both implementations can be constrained to match a known mean.
\end{abstract}

\noindent%
\vfill

\newpage
\spacingset{1.45} 
\section{Introduction}
\label{sec:intro}

Surveys often ask respondents to report income in brackets or \textit{bins}, such as \$0-10,000, \$10,000-20,000,\ldots,\$200,000+. Even in surveys where respondents report exact incomes, incomes may be binned before publication, either to protect privacy or to summarize the income distribution compactly with the number of incomes in each bin. Table \ref{tab:nantucket} gives a binned summary of household incomes in Nantucket, the richest county in the US. 

\begin{table}[tbh]
\centering
\begin{threeparttable}
\caption{Household incomes in Nantucket}
\label{tab:nantucket}
\begin{tabular}{cccc}
\hline
{\ul Min} & {\ul Max} & {\ul Households} & {\ul Cumulative distribution} \\
\$0         & \$10,000  & 165              & 5\%                 \\
\$10,000  & \$15,000  & 109              & 8\%                 \\
\$15,000  & \$20,000  & 67               & 9\%                 \\
\$20,000  & \$25,000  & 147              & 13\%                \\
\$25,000  & \$30,000  & 114              & 17\%                \\
\$30,000  & \$35,000  & 91               & 19\%                \\
\$35,000  & \$40,000  & 148              & 23\%                \\
\$40,000  & \$45,000  & 44               & 24\%                \\
\$45,000  & \$50,000  & 121              & 28\%                \\
\$50,000  & \$60,000  & 159              & 32\%                \\
\$60,000  & \$75,000  & 358              & 42\%                \\
\$75,000  & \$100,000 & 625              & 59\%                \\
\$100,000 & \$125,000 & 338              & 69\%                \\
\$125,000 & \$150,000 & 416              & 80\%                \\
\$150,000 & \$200,000 & 200              & 86\%                \\
\$200,000 &           & 521              & 100\%    \\ \hline
\end{tabular}
 \begin{tablenotes}
    \small
    \item
      \textit{Note}. Each bin's population is estimated from a 1-in-8 sample of households who took the American Community Survey in 2006-10. Incomes are in 2010 dollars.
    \end{tablenotes}

\end{threeparttable}
\end{table}

Binning presents challenges to investigators who want to estimate simple summary statistics such as the mean, median, or standard deviation, or inequality statistics such as the Gini coefficient, the Theil index, the coefficient of variation, or the mean log deviation. Researchers have implemented several methods for calculating estimates from binned incomes.

The simplest and most popular approach is to assign each case to the midpoint of its bin---using a robust pseudo-midpoint for the top bin, whose upper bound is typically undefined  (e.g., Table \ref{tab:nantucket}). The weakness of the midpoint approach is that it treats income as a discrete variable, but the method also has several strengths. The midpoint method is easy to implement and runs quickly. Midpoint estimates are also ``bin-consistent'' \citep{vonHippel2015} in the sense that midpoint estimates will get arbitrarily close to their estimands if the bins are sufficiently numerous and narrow. 

Another approach is to fit the bin counts to a continuous parametric distribution. Popular distributions include 2-, 3-, and 4-parameter distributions from the generalized beta family, which includes the Pareto, lognormal, Weibull, Dagum, and other distributions \citep{McDonald2008}. One implementation fits up to 10 distributions and selects the one that fits best. An alternative is to use the AIC or BIC to calculated a weighted average of income statistics across several candidate distributions \citep{vonHippel2015}.

A strength of the parametric approach is that it treats income as continuous. A weakness is that even the best-fitting parametric distribution may not fit the bin counts particularly well. If the fit is poor, the parametric approach is not bin-consistent; that is, even with an infinite number of bins, each infinitesimally narrow, a parametric distribution may produce poor estimates if it is not a good fit to the underlying distribution of income. 

A practical weakness of the parametric approach is that it is typically implemented using iterative methods which can be slow. The speed of a parametric fit may be acceptable if you fit a single distribution to a single binned dataset, but runtimes of hours are possible if you fit several distributions to thousands of binned datasets---such as every county or school district in the US. Other computational issues include nonconvergence and undefined estimates. These issues are rare but inevitable when you run thousands of binned datasets \citep{vonHippel2015}.

Neither approach---midpoint or parametric---is uniformly more accurate. With many bins, midpoint estimates are better because of bin-consistency, while with fewer than 8 bins, parametric estimates can be better because of their smoothness \citep{vonHippel2015}. Empirically, the parametric and midpoint approaches produce similarly accurate estimates from typical US income data with 15 to 25 bins. Both methods typically estimate the Gini within a few percentage points of its true value. This is accurate enough for many purposes, but can lead to errors when estimating small differences or changes such as the 5\% increase in the Gini of US family income that occurred between 1970 and 1980 \citep{vonHippel2015}. 

A potential improvement is to fit binned incomes to a flexible nonparametric continuous density. Like parametric densities, a nonparametric density treats income as continuous. Like the midpoint method, a nonparametric density can be bin-consistent and fit the bin counts as closely as we like.  

Unfortunately, past nonparametric approaches have been disappointing. A nonparametric approach using kernel density estimation had substantial bias under some circumstances \citep{Minoiu2012}. A nonparametric approach using a spline to model the log of the density \citep{Kooperberg1992} had even greater bias \citep{vonHippel2012}, though it was not clear whether the bias came from the method or its software implementation.

In this paper, we implement and test a nonparametric continuous method that outperforms its predecessors in both speed and accuracy. The method, which we call \textit{CDF interpolation}, simply connects points on the empirical cumulative distribution function (CDF). The method can connect the points using line segments or cubic splines. When cubic splines are used, the method is similar to ``histospline'' or ``histopolation'' methods which fit a spline to a histogram \citep{Wahba1976,Morandi1989}. But histosplines are limited to histograms which have bins of equal width \citep{Wang2015}. CDF interpolation is more general approach which can handle income data where the bins have unequal width and the top bins has no upper bound (e.g., Table \ref{tab:nantucket}).

We have implemented CDF interpolation in our R package \textit{binsmooth} \citep{Hunter2016}, which is available for download from the Comprehensive R Archive Network (CRAN).  

Our results will show that statistics estimated with CDF interpolation are slightly more accurate than estimates obtained using midpoints or parametric distributions. In addition, CDF interpolation is much faster than parametric estimation, though not as fast as the midpoint method.

We also show that the differences between methods are dwarfed by the improvement we get if we constrain a method to match the grand mean of income, which the US Census Bureau often reports alongside the bin counts. If we constrain either the interpolated CDF or the midpoint method to match a known mean, we get dramatically better estimates of the Gini. Our \textit{binsmooth} package can constrain an interpolated CDF to match a known mean, and our new version (2.0) of the \textit{rpme} command for Stata can constrain the midpoint method to match a known mean as well.

In the rest of this paper, we define the midpoint, parametric, and interpolated CDF methods more precisely, then compare the accuracy of estimates in binned data summarizing household incomes within US counties. We also show how much estimates improve if we have the mean as well as the bin counts. 

\section{Methods}
\label{sec:meth}

\subsection{Binned data}
A binned data set, such as Table \ref{tab:nantucket}, consists of counts $n_1, n_2, \ldots, n_B$ specifying the number of cases in each of $B$ bins. The total number of cases is $T = n_1 + n_2 + \cdots + n_B $. Each bin $b$ is defined by an interval $[l_b,u_b), b=1, \ldots, B$, where $l_b$ and $u_b$ are the lower and upper bound of income for that bin. The bottom bin often starts at zero ($l_1=0$), and the top bin may have no upper bound ($u_B=\infty$). 

\subsection{A midpoint method}

The oldest and simplest way to analyze binned data is the midpoint method, which within each bin $b$ assigns incomes to the bin midpoint $m_b=(l_b+u_b)/2$ (\citealt{}{Heitjan1989} for a review). Then statistics such as the Gini can be calculated by applying sample formulas to the midpoints $m_b$ weighted by the counts $n_b$.

When the top bin has no upper bound, we must define a pseudo-midpoint for it. The traditional choice is ${\mu}_B = l_B \alpha/(\alpha-1)$, which would define the arithmetic mean of the top bin if top-bin incomes followed a Pareto distribution with shape parameter $\alpha > 1$ \citep{Henson1967}. The problem with this choice is that the arithmetic mean of a Pareto distribution is undefined if $\alpha \leq 1$ and grows arbitrarily large as $\alpha$ approaches 1 from above. 

A more robust choice is the harmonic mean $h_B = l_B (1+1/\alpha)$, which is defined for all $\alpha>0$ \citep{vonHippel2015}. We use the harmonic mean in this article. We estimate $\alpha$ by fitting a Pareto distribution to the top two bins and calculating the maximum likelihood estimate from the following formula \citep{Quandt1966}:
\begin{equation}
    \hat\alpha = \frac{ln((n_{B-1}+n_B)/n_B)}
                        {ln(l_B/l_{B-1})}
    \label{eqn:alpha}
\end{equation}

If the survey provides the grand mean $\mu$, then we need not assume that top-bin incomes follow a Pareto distribution. Instead, we can calculate 
\begin{equation}
    \hat\mu_B = \frac{1}{n} (T \mu - \sum_{b=1}^{B-1} n_B m_B )
    \label{eqn:top_bin_mean}
\end{equation}

\noindent which would be the mean of the top bin if the means of the lower bins were the midpoints $m_b$. Then $\hat\mu_B$ can serve as a pseudo-midpoint for the top bin. 

When estimated in this way, it occasionally happens (e.g., in 4 percent of US counties) that the top bin's mean $\hat\mu_B$ is slightly less than its lower bound $l_B$. This is infelicitous, but $\hat\mu_B$ can still be used, and the resulting Gini estimates are not necessarily bad.\footnote{It might be a little better to set $\hat\mu_B=l_B$ and move the other midpoints slightly the the left, but this would affect only 4 percent of counties, and those only slightly since the affected counties typically have few cases in the top bin.}

The midpoint methods described in this section are implemented by the \textit{rpme} command for Stata \citep{Duan2016}, where  \textit{rpme} stands for ``robust Pareto midpoint estimator'' \citep{vonHippel2016}. Except for mean-matching, the approach is also implemented in the \textit{binequality} package for R \citep{Scarpino2017}.

\subsection{Fitting parametric distributions}

The weakness of the midpoint method is that it treats income as discrete. An alternative is to model income as a continuous variable $X$ that fits some parametric CDF $F(X|\theta)$. Here $\theta$ is a vector of parameters, which can be estimated by iteratively maximizing the log likelihood:
\begin{equation}
    \ell(\theta|X) = ln \prod_{b=1}^{B} (P(l_b<X<u_b))^{n_b}
        \label{eqn:log_likelihood}
\end{equation}

\noindent where $P(l_b<X<u_b)=F(u_b)-F(l_b)$ is the probability, according to the fitted distribution, that an income is in the bin $[l_b,u_b)$.\footnote{This formula includes the top bin $B$, where $u_B=\infty$ and $F(u_B)=1$. Some older articles (e.g., \citep{McDonald1979,McDonald1995,Bandourian2002}) add the following constant to the log likelihood: $ln(T!)-\sum_{b=1}^{B}{ln(n_b!)}$. This is not wrong, but it is unnecessary since adding a constant does not change the parameter values at which the log likelihood is maximized \citep{edwards1972}.}

While any parametric distribution can be considered, in practice it is hard to fit a distribution unless the number of parameters is small compared to the number of well-populated bins. Most investigators favor 2-, 3-, and 4-parameter distributions from the generalized beta family \citep{McDonald2008}, which includes the following 10 distributions: the log normal, the log logistic, the Pareto (type 2), the gamma and generalized gamma, the beta 2 and generalized beta (type 2), the Dagum, the Singh-Maddala, and the Weibull. 

\textit{A priori} it is hard to know which distribution, if any, will fit well. The fit of a distribution can be tested by the following goodness-of-fit likelihood ratio statistic \citep{vonHippel2015}:

\begin{equation}
G^2=-2 (\hat{\ell} - \sum_{b=1}^{B}{n_b ln(n_b/T)}
\end{equation}

\noindent where $\hat{\ell}$ is the maximized log likelihood. Under the null hypothesis that the fitted distribution is the true distribution of income, $G^2$ would follow a chi-square distribution with $B^*-k$ degrees of freedom, where $k$ is the number of parameters, $B^*=min(B_{>0},B-1)$, and $B_{>0}$ is the number of bins with nonzero counts. We reject the fit of a distribution if $G^2$ has $p<.05$ in the null distribution. 

In empirical data it is common to reject every distribution in the generalized beta family \citep{Bandourian2002,vonHippel2015}. In addition, some distributions may fail to converge, or may converge on parameter values that imply that the mean or variance of the distribution is undefined \citep{vonHippel2015}. 

A solution is to fit all 10 distributions, screen out any with undefined moments and, among the distributions remaining, select the one that fits best according to the Akaike or Bayes information criterion (AIC or BIC):
\begin{align*}
AIC &= 2k - 2 \hat{\ell} \\
BIC &= ln(T) k  - 2 \hat{\ell}
\end{align*}

Or, instead of selecting a single best-fitting distribution, one can average estimates across several candidate distributions weighted proportionately to a function of the AIC or BIC (specifically $exp(-AIC/2)$ or $exp(-BIC/2)$) an approach known as \textit{model averaging}  \citep{Burnham2004}. In general, model averaging yields better estimates than model selection, but when modeling binned incomes the advantage of model averaging is negligible \citep{vonHippel2015}. 

Although it sounds broad-minded to fit 10 different distributions, there is limited diversity in the generalized beta family. All the distributions in the generalized beta family are unimodal and skewed to the right. Some distributions are quite similar (e.g., Dagum and generalized beta), and others rarely fit well (e.g., log normal, log logistic, Pareto). So in practice the range of viable and contrasting distributions in the generalized beta family is small; you can fit just 3 well-chosen distributions (e.g., Dagum, gamma, and generalized gamma) and get estimates almost as good as those obtained from fitting all 10 \citep{vonHippel2015}.  

We use the fitted distributios to estimate income statistics such as the mean, variance, Gini, or Theil. Sometimes the income statistic is a simple function of the distributional parameters $\theta$, but other times the function is unknown, or hard to calculate. As a general solution, it is easier to calculate income statistics by applying numeric integration to appropriate functions of the fitted distribution \citep{McDonald2008}.

When the grand mean is available, the distributional parameters could in theory be constrained to match the grand mean as well as approximate the bin counts. This would be difficult, though, since for distributions in the generalized beta family the mean is a complicated nonlinear function of the parameters. We have not attempted to constrain our parametric distributions to match a known mean.

The parametric approaches described in this section are implemented in the \textit{binequality} package for R \citep{Scarpino2017} and the \textit{mgbe} command for Stata \citep{Duan2016}. Here \textit{mgbe} stands for ``multi-model generalized beta estimator'' \citep{vonHippel2015}.

\subsection{Interpolated CDFs}

Since parametric distributions may fit poorly, an alternative is to define a flexible nonparametric density that fits the bin counts exactly. Our nonparametric approach is called CDF interpolation.

To understand CDF interpolation, consider that binned data define $B$ discrete points on the empirical cumulative distribution function (CDF). The empirical CDF for Nantucket is given in the last column of Table \ref{tab:nantucket}, which shows that 5\% of Nantucket households make less than \$10,000, 8\% make less than \$15,000, and so on. Formally, at an income of 0 the empirical CDF is $\hat{F}(0)=0$, and at each bin's upper bound $u_b$ the empirical CDF is the observed fraction of incomes that are less than $u_b$---i.e., $\hat{F}(u_b) = (n_1 + n_2 + \cdots + n_b)/T$.\footnote{If the binned incomes come from a sample (as they do in Table \ref{tab:nantucket}), then the estimate $\hat{F}(u_b)$ may differ from the true CDF ${F}(u_b)$ because of sampling error.}

Now to estimate a continuous CDF $\hat{F}(x), x>0$, we just connect the dots. That is, we define a continuous nondecreasing function that interpolates between the $B$ discrete points of the empirical CDF. 

Then the estimated probability density function (PDF) is just the derivative of the interpolated CDF $\hat{F}(x)$. Note that the estimated PDF ``preserves areas'' \citep{Morandi1989}---i.e., preserves bin counts. That is, within in each bin ($\hat{P}(l_b<x<u_b)=\hat{F}(u_b)-\hat{F}(l_b)$), the total of the estimated density is equal to the observed fraction of incomes that are in that bin ($n_b/T$). 

The shape of the PDF depends on the function that interpolates the CDF. 

\begin{itemize}
 \item If the CDF is interpolated by line segments, then the CDF is polygonal, and the PDF is a step function that is discontinuous at the bin boundaries.
 \item If the CDF is interpolated more smoothly, by a continuously differentiable monotone cubic spline, then the PDF is piecewise quadratic --- i.e., continuous at the bin boundaries and quadratic between them.
 \end{itemize}

There remains a question of how to shape the CDF in the top bin, which typically has no upper bound. In our implementation, the CDF of the top bin can be rectangular, exponential, or Pareto.\footnote{In our implementation, the exponential and Pareto tails are approximated by a sequence of rectangles of decreasing heights.} Each of these distributions has one parameter, which is estimated as follows.

\begin{itemize}
\item If the grand mean of income is known, then the parameter shaping the top bin is constrained so that the mean of the fitted distribution matches the grand mean. It occasionally happens (e.g., in 4 percent of US counties) that we cannot reproduce the known mean this way, because the known mean is already less than the mean of the lower $B-1$ bins without the tail. In that case, we make an \textit{ad hoc} adjustment by shrinking the bin boundaries toward the origin -- that is, by replacing $(l_b,u_b)$ with $(s l_b,s u_b)$, where the shrinkage factor $s<1$ is chosen so that a small tail can be added to reproduce the grand mean. The shrinkage factor is rarely less than .995.

\item If the mean income is not known, we substitute an \textit{ad hoc} estimate. We obtain that estimate by temporarily setting the upper bound of the top bin to $u_B=2l_B$ and calculating the mean of a step PDF fit to all $B$ bins. Then we unbound the top bin and proceed as though the mean were known. 
\end{itemize}

Income statistics are estimated by applying numerical integration to functions of the fitted PDF or CDF.  



The methods in this section are implemented by our \textit{binsmooth} package for R \citep{Hunter2016}. Within the \textit{binsmooth} package, the \textit{stepbins} function implements a step-function PDF (and polygonal CDF), while the \textit{splinebins} function implements a cubic spline CDF (and piecewise quadratic PDF).


\subsection{Recursive subdivision}

Another way to obtain a smooth PDF estimate that preserves bin areas is to subdivide the bins into smaller bins, and then adjust the heights of the subdivided bins to shorten the jumps at the bin boundaries. This method, \textit{recursive subdivision}, is implemented by the \textit{recbin} command in the \textit{binsmooth} package for R \citep{Hunter2016}. Recursive subdivision is slower and more computationally intensive than CDF interpolation and the resulting estimates are practically identical. We present the details of recursive subdivision in the Appendix.

\section{Data and Results}

Between 2006 and 2010, the American Community Survey (ACS) took a 1-in-8 sample of US households \citepalias{uscacs}. Household incomes were inflated to 2010 dollars and summarized in binned income tables for  each of the 3,221 US counties. The published bin counts are estimates of the population counts. We can approximate the sample counts by dividing the population counts by 8. Dividing counts by a constant makes no difference to any of our statistics, except for the BIC and $G^2$ statistics that are used when fitting parametric distributions.

The Census also published means and Gini coefficients for each county \citep{Bee2012}. These statistics were estimated from exact incomes before binning, and so are more accurate than any estimate that could be calculated from the binned data. They are sample estimates which may differ from population values, but they remain a useful standard of comparison for our binned-data estimates.

\subsection{Results for Nantucket}

\begin{figure}[hbt]
        \centering
         \caption{Different PDFs for the Nantucket data.}
        \includegraphics[width=6.5in]{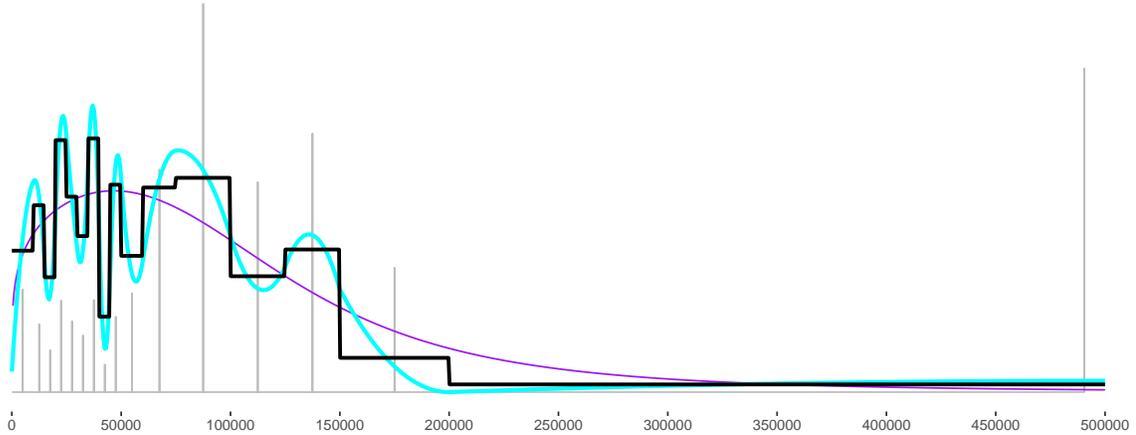}
        \label{fig:comparepdfs}
        \floatfoot{The Dagum distribution is drawn in purple, the quadratic spline in blue, and the step function in black. Gray spikes illustrate the midpoint method.}
    \end{figure}

Table \ref{tab:nantucket} summarized the binned incomes for Nantucket County. Figure~\ref{fig:comparepdfs} fits several distributions to the Nantucket data. 

The midpoint method is illustrated by gray spikes at the bin midpoints; the spike heights are proportional to the bin counts. The black step function is the PDF implied by a linear interpolation of the CDF, and the blue curve is the piecewise quadratic PDF implied by a cubic spline of the CDF. Both CDF interpolations fit the bin counts perfectly; in fact, their jagged appearance suggests they may \textit{overfit} the data---a concern that we will revisit in the Conclusion. The step PDF looks slightly less volatile than the piecewise quadratic PDF, suggesting that the step PDF may be less overfit.

The purple curve is the Dagum distribution. The Dagum fits Nantucket better than other distributions from the generalized beta family, but it does not fit well. It fails the $G^2$ goodness-of-fit test, and visually it fits the bin counts poorly. For example, between \$70,000 and \$150,000 the Dagum curve suggests there should be substantially fewer households than there are, and above \$150,000 it suggests that there should be more.

Except for the Dagum distribution, all the methods in Figure~\ref{fig:comparepdfs} are calibrated to reproduce the grand mean.
    
Table~\ref{tab:nantucket_estimates} summarizes the Nantucket estimates. The true mean is \$137,000 and the true Gini is .547. All the methods underestimate these quantities. When fit without knowledge of the true mean, every method underestimates the mean by 12 to 20 percent and the Gini by 15 to 21 percent, with the simple midpoint method coming closer than its more sophisticated competitors.

\begin{table}[hbt]
\centering
\caption{Estimates for Nantucket}
\label{tab:nantucket_estimates}
\begin{tabular}{llllll}
\hline
                   & {\ul Estimate}      & {\ul Mean}      & {\ul Gini} \\
                   & True                     & \$137,811 & .547 &  &  \\
Without the true mean  & Midpoint (RPME)          & \$121,506 & .464 &  &  \\
                   & Parametric (MGBE)        & \$112,960 & .453 &  &  \\
                   & CDF interpolation (linear) & \$110,419 & .438 &  &  \\
                   & CDF interpolation (cubic spline)  & \$110,419 & .433 &  &  \\
Matching the true mean & Midpoint                 &           & .510 &  &  \\
                   & CDF interpolation (linear) &           & .537 &  &  \\
                   & CDF interpolation (cubic spline)  &           & .525 &  & \\
                   \hline
\end{tabular}
\end{table}

When given the true mean, the midpoint and CDF interpolation methods do much better. They still underestimate the Gini, but only by 2 to 7 percent. The closest estimate is obtained by linear interpolation of the CDF. A smoother cubic spline interpolation does a little worse, but still better than the midpoint method.

Although the estimates for Nantucket are less accurate overall than the estimates for most other counties, the relative performance of different methods in Nantucket is similar to what we will see elsewhere.  

\subsection{Results for all US counties}
\label{sec:bakeoff}

\begin{figure}[hbt]
        \centering
        \includegraphics[width=6in]{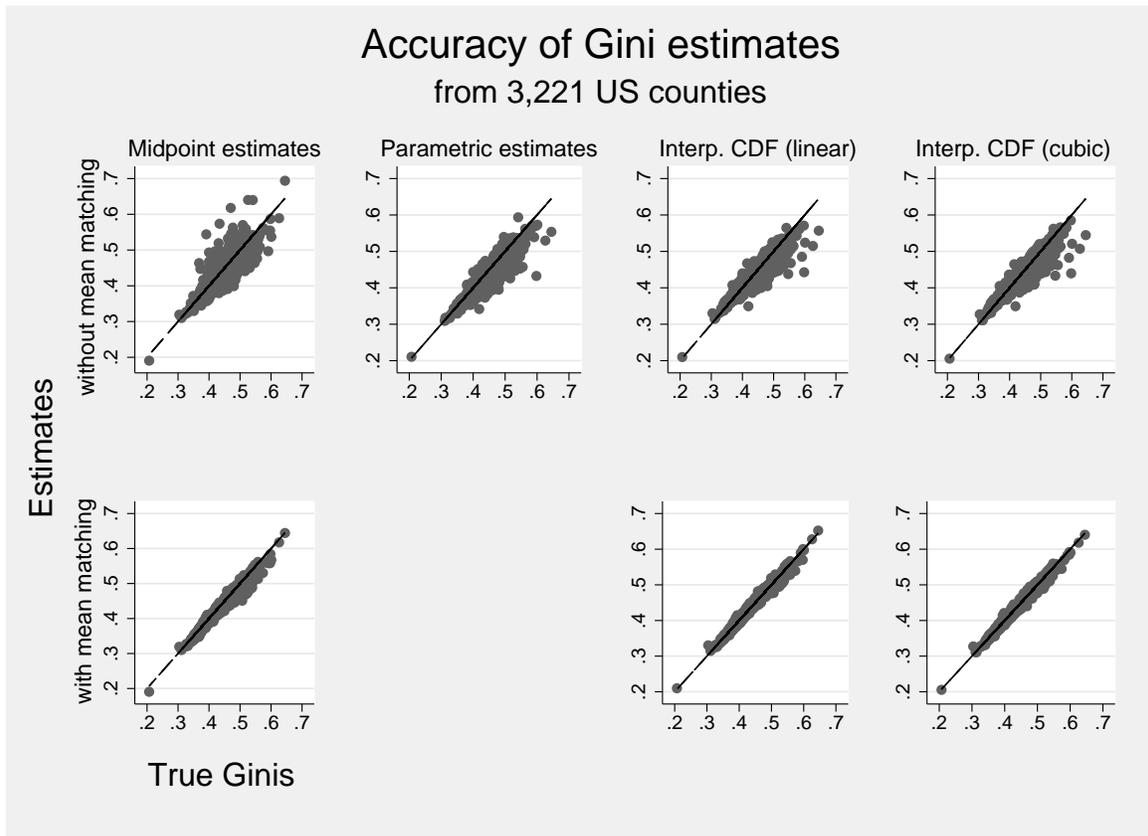}
        \caption{Accuracy of Gini coefficients estimated by different methods, with and without mean-matching.}
        \label{fig:scatterplots}
    \end{figure}

Figure~\ref{fig:scatterplots} evaluates all county Gini estimates graphically by plotting the estimated Gini $\hat{\theta}_j$ of each county against the published Gini ${\theta}_j$. In the bottom row, where the methods are constrained to match the published mean, the Gini estimates are close to a diagonal reference line ($\hat{\theta}_j=\theta$) indicating nearly perfect estimation. In the top row, where the methods do not match the mean, the estimates are more scattered, indicating lower accuracy.

We can summarize the accuracy of Gini estimates in several ways. For a single county $j$, the percent estimation error is $e_j = 100\times(\hat{\theta}_j - \theta_j)/\theta_j$. Then across all counties, the \textit{percent relative bias} is the mean of $e_j$, the \textit{percent root mean squared error} (RMSE) is the square root of the mean of $e_j^2$, and the \textit{reliability} is the squared correlation between $\theta_j$ and $\hat{\theta}_j$. 

Table~\ref{tab:errorres} summarizes our findings. If the estimators ignore the published county means, then estimated Ginis have biases between 0\% and -3\%, RMSEs between 3\% and 4\%, and reliabilities between 82\% and 88\%. The interpolated CDF estimates have the best bias, the best RMSE, and the second best reliability, and they are just as good with linear interpolation as with cubic spline interpolation. The parametric estimates have the best reliability, but the worst bias, the worst RMSE, and by far the worst runtime, at 4.5 hours. 

\begin{table}[tbh]
\begin{threeparttable}[tbh]
    \caption{Speed and accuracy of Gini estimates for 3,221 US counties.  
    \label{tab:errorres}}
\centering
\begin{tabular}{llcccl}
\hline
                  &     {\ul Estimator}              & {\ul Bias} & {\ul RMSE} & {\ul Reliability} & {\ul Runtime} \\
Without true mean & Midpoint                        & -2\%       & 4\%            & 82\%              & 4 sec         \\
                  & Parametric                      & -3\%       & 4\%            & 89\%              & 4.5 hr        \\
                  & CDF interpolation (linear)     & 0\%        & 3\%            & 88\%              & 40 sec        \\
                  & CDF interpolation (cubic spline) & -1\%       & 3\%            & 88\%              & 2 min         \\
Matching true mean    & Midpoint                               & -1\%       & 2\%            & 98\%              & 7 sec         \\
                  & CDF interpolation (linear)     & 0\%        & 1\%            & 99\%              & 36 sec        \\
                  & CDF interpolation (cubic spline) & -1\%       & 1\%            & 99\%              & 2 min      \\
\hline                  

\end{tabular}
    \begin{tablenotes}
    \small
    \item
\textit{Note}. RMSE=root mean squared error.
Runtimes on an Intel i7 Core processor with a speed of 3.6 GHz.
    \end{tablenotes}
    \end{threeparttable}
\end{table}

When the methods are constrained to match the published county means, the estimates improve dramatically. The bias shrinks to 0-1\%, the RMSE shrinks to 1-2\%, and the reliability grows to 98-99\%. The midpoint estimates are excellent, and the interpolated CDF estimates are even better, and just as good with linear interpolation as with cubic spline interpolation.

The differences among the methods are much smaller than the improvement that comes from constraining any method to match the mean. Of course, this observation is only helpful when the mean is known.

\section{Conclusion}
\label{sec:conc}

CDF interpolation produces estimates that are at least a little better than midpoint or parametric estimates, whether the true mean is known or not. And CDF interpolation runs much faster than parametric estimation, thought not as fast as midpoint estimation.

We initially suspected that cubic spline interpolation would improve on simple linear interpolation, but empirically this turns out to be false. In estimating county Ginis, linear CDF interpolation was at least as accurate as cubic spline interpolation. 

The accuracy of linear CDF interpolation is remarkable, since it implies a step function for the PDF. Step PDFs look clearly unrealistic, especially in the top and bottom bins where the step function is flat while the true distribution likely has an upward or downward slope \citep{cloutier1995}. Our step PDF permits a downward Pareto or exponential slope in the top bin, and our cubic spline CDF can fit an upward slope to the bottom bin. But neither of these refinements does much to improve the accuracy of Gini estimates.

The differences in accuracy among the methods are small, and they are dwarfed by the improvement in accuracy that comes from knowing the grand mean. By constraining binned-data methods to match a known mean, we can typically get county Gini estimates that are within 1-2\% of the estimates we would get if the data were not binned. Our \textit{binsmooth} package for R can constrain interpolated CDFs to match a known mean, and our \textit{rpme} command for Stata can constrain the top-bin midpoint to match a known mean as well. We have not constrained our parametric distributions to match a known mean; we believe it would be difficult to do so.

While the mean-constrained estimates are very accurate, there may be room for improvement when the mean is unknown. Perhaps the most promising idea is additional smoothing. As we noticed in Figure~{fig:comparepdfs}, interpolated CDFs can be a bit jagged and may ``overfit'' the sample in the sense that they find nooks and crannies that might not appear in another sample or in the population. Likewise interpolated CDFs may be overfit to a specific set of bin boundaries. If the fitted CDF were a little smoother and did not quite preserve the counts of the least populous bins, it might fit the population and other samples (perhaps with different bin boundaries) a little better.

    \bibliography{binsmoothing}

\appendix

\section{Appendix: PDF smoothing by recursive subdivision}
\label{apx:recsub}

Recursive subdivision is another way to smooth the fitted PDF. Like CDF interpolation, recursive subdivision preserves bin areas, but recursive subdivision is more computationally intensive and produces very similar results. Recursive subdivision is implemented by the \textit{recbins} function in our \textit{binsmooth} package for R.

A slight change of notation will be helpful. Since the upper bound $u_b$ of each bin is equal to the lower bound $l_{b+1}$ of the next, we can think the bins as having a set of ``edges'' ${e_0,e_1,\ldots,e_{B}}$, where $e_0=0$, and the other $e_b=u_b$ are the upper bounds of bins $1,\ldots,B$. 

\begin{figure}[ht]
        \centering
        \includegraphics[width=5in]{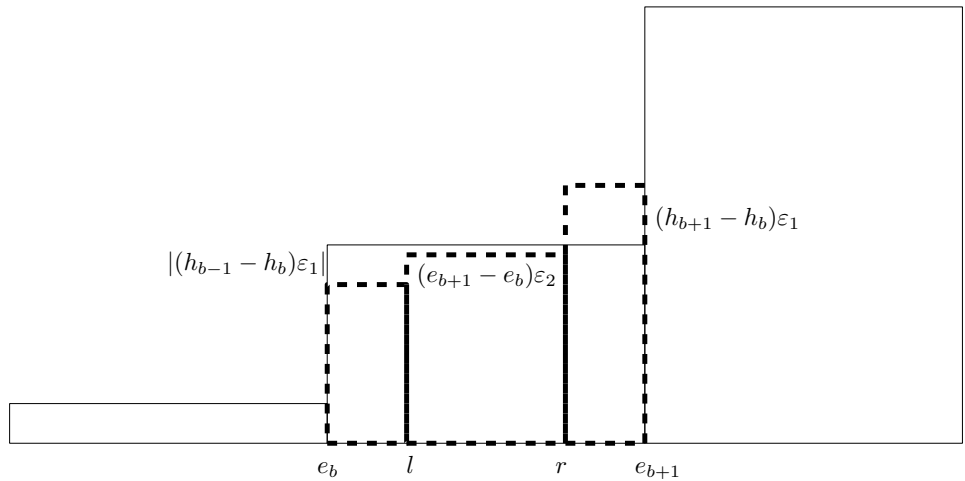}
        \caption{Bin subdivision. Each original bin is replaced by three new bins (bold, dashed) such that the bin area is preserved.}
        \label{fig:recsub}
    \end{figure}

Start by fitting a step PDF. Let $h_b$ be the height of the step PDF in the bin $[e_b, e_{b+1})$. Given parameters $\varepsilon_1 \in (0,0.5)$ and $\varepsilon_2 \in (0,1)$, the subdivision process begins by introducing new bin edges $l$ and $r$ between $e_b$ and $e_{b+1}$ such that $(l+r)/2 = (e_b+e_{b+1})/2$ and $r-l = (e_{b+1}-e_b)\varepsilon_2$. The height of the new bin on the left with edges $e_b$ and $l$ is then shifted horizontally by $(h_{b-1}-h_b)\varepsilon_1$, while the height of the new bin on the right with edges $r$ and $e_{b+1}$ is shifted horizontally by $(h_{b+1}-h_b)\varepsilon_1$. Finally, the new middle bin with edges $l$ and $r$ is shifted horizontally so that the area of the three new bins equals the area of the original bin.\footnote{We do not subdivide the bin in those rare cases where subdivision would yield a middle bin with negative height.} See Figure~\ref{fig:recsub}.

In order for the above formulas to apply to the top and bottom bins, we create pseudo-bins above and below them with heights of zero. This ensures that the subdivided PDF will tend toward a height of zero at the lower edge of the bottom bin and the upper edge of the top bin.

\begin{figure}[hbt]
        \centering
        \includegraphics[width=6in]{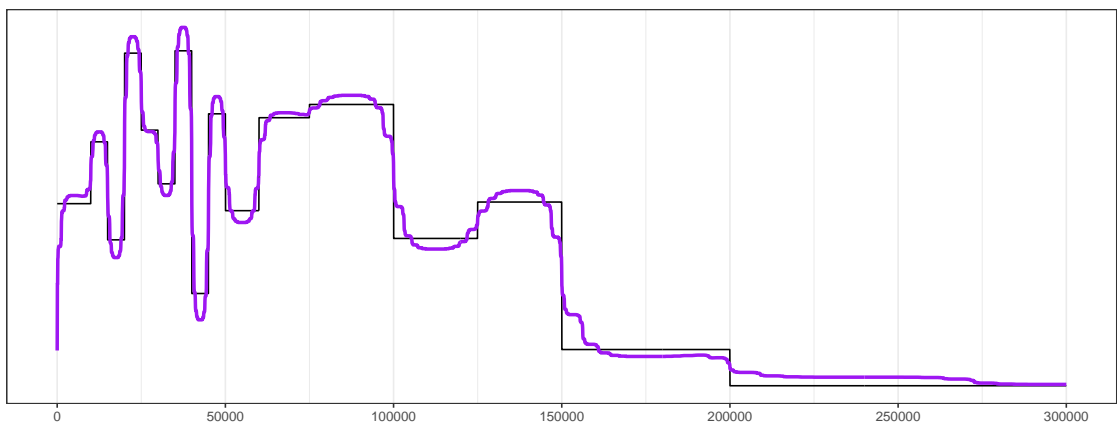}
        \caption{Recursively subdivided density for the Nantucket data (purple). A step function (black) is shown in the background.}
        \label{fig:recsubnan}
    \end{figure}

     The smoothed PDF is obtained from the step PDF by applying the subdivision process to each bin, then applying the process again to each subdivided bin, and so on, until the desired level of smoothness is reached. In practice, three rounds of subdivision are sufficient to produce a reasonably smooth PDF, and we found that choosing $\varepsilon_1 = 0.25$ and $\varepsilon_2 = 0.75$ produced nicely smoothed PDF's from most empirical data sets. Figure~\ref{fig:recsubnan} shows the result of recursive subdivision in Nantucket. 
     
     Unfortunately, if the original step PDF was constrained to match a known mean, the subdivision process may cause the mean to deviate slightly. But the estimated Gini typically remains quite accurate.
     
\end{document}